\documentclass[10pt, conference, compsocconf]{IEEEtran}

\usepackage{caption} 
\usepackage{booktabs}

\newcommand\blfootnote[1]{%
  \begingroup%
  \renewcommand\thefootnote{}\footnote{#1}%
  \addtocounter{footnote}{-1}%
  \endgroup%
}

\usepackage{multirow}
\usepackage{balance}
\usepackage{nohyperref}
\usepackage{url}
\usepackage{color}
\usepackage{todonotes}
\usepackage{siunitx}
\usepackage{svg}

\usepackage[stable]{footmisc}

\begin{document}\sloppy


\title{Universal EEG Encoder for Learning Diverse Intelligent Tasks}







\author{\IEEEauthorblockN{Baani Leen Kaur Jolly*, Palash Aggrawal*, Surabhi S Nath*, Viresh Gupta*,  Manraj Singh Grover, Rajiv Ratn Shah}
\IEEEauthorblockA{MIDAS Lab, IIIT-Delhi\\
\{baani16234, palash16064, surabhi16271, viresh16118, manrajg, rajivratn\}@iiitd.ac.in}
}

\maketitle%
\blfootnote{* Student authors contributed equally to this work}%
\begin{abstract}
Brain Computer Interfaces (BCI) have become very popular with Electroencephalography (EEG) being one of the most commonly used signal acquisition techniques. A major challenge in BCI studies is the individualistic analysis required for each task. Thus, task-specific feature extraction and classification are performed, which fails to generalize to other tasks with similar time-series EEG input data. To this end, we design a GRU-based universal deep encoding architecture to extract meaningful features from publicly available datasets for five diverse EEG-based classification tasks. Our network can generate task and format-independent data representation and outperform the state of the art EEGNet architecture on most experiments. We also compare our results with CNN-based, and Autoencoder networks, in turn performing local, spatial, temporal and unsupervised analysis on the data.
\end{abstract}
%

\section{Introduction}
Brain Computer Interfaces (BCI) create a direct communication of the human brain with an external agent and serves applications in the fields of healthcare, neuroprosthetics, rehabilitation, robotics, entertainment, security among many others \cite{BCI-VR-IoT, BCI-Military}.
Electroencephalogram (EEG) based signal acquisition is a non-invasive technique \cite{non_invasive} wherein electrodes on the scalp record signals with respect to time. Analyzing EEG signals requires effective methods to extract meaningful features which are usually tailored to a particular task.



A universal encoding can allow efficient comparison and evaluation across multiple datasets using a shared network for generating the feature vector. A common feature encoding can help in studying the patterns and relations between various types of EEG signals. This can potentially eradicate the need for task dependent investigation and manual handcrafted feature extraction which is often very cumbersome.

Recent work by Lawhern et al. \cite{Lawhern2018} 
proposed EEGNet, which develops a compact convolutional neural network (CNN) for EEG data. This network performs consistently well across different downstream tasks. A major limitation of EEGNet is that it uses a CNN network reflecting its inability to capture the sequential nature of the data.

In this paper, we propose a GRU-based universal deep learning encoding architecture for various EEG-based downstream tasks. The GRU (Gated Recurrent Unit) \cite{learning-rnn-encoder-cho2014, properties-neural-machine-translation-cho2014} incorporates temporal dependencies in the data and can thus overcome the limitations of EEGNet. We also compare its performance with other networks like CNN-based network \cite{Tirupattur2018} and Autoencoder networks (Section \ref{methodology}), on five significantly different EEG-based datasets including Emotion Detection \cite{seed_original}, Digit Recognition (Section \ref{MNIST_Brain}), Object Recognition \cite{Tirupattur2018}, Task Identification \cite{bci_smr} and Error Detection \cite{ern_kaggle}. The proposed universal encoding architecture yields comparable results and in many cases outperforms the state of the art methods across all downstream tasks.

\section{Literature Survey}\label{Lit_survey}
EEG signals are one of the most common methods of mapping brain activity due to its non-invasive nature, portability, cost, safety, and high temporal resolution. Other methods include Positron Emission Tomography (PET) and Functional Magnetic Resonance Imaging (fMRI) have a lower time resolution and unlike EEG, measure changes in blood flow or metabolic activities, which are indirect indicators of brain electrical activity. 

Previous work in this field have used EEG signals to classify hand motions \cite{6614007}, songs based on human thoughts \cite{Bauer2015}, for medical purposes such as diagnosing epilepsy \cite{8389880} and Autism Spectrum Disorder \cite{autismarticle}, rehabilitation purposes \cite{motorControl}, and as feedback mechanisms in therapies for Post-Traumatic Stress Disorder \cite{VRStress}, Attention-Deficit Hyperactivity Disorder \cite{VRBiofeedback} and several other applications.

Since EEG signals are time series data, extracted features (e.g. Wavelet Transformation, Independent Component Analysis, Autoregression, and Empirical Mode Decomposition \cite{reviewPaperEEGExtraction}) must represent the data meaningfully. 
However, these 
are manually engineered making it difficult to select the best set of extraction methods and wave properties to capture the complex nature of EEG data. Previous work suggests eigenfeature methods (like computing Power Spectral Densities on segmented EEG \cite{EigenSignalAnalysis}), Fast Fourier Transform \cite{8614866} and Genetic Algorithms \cite{Wen2017}.

With the rise in the popularity of Deep Learning methods, the task of extracting optimal features from EEG data has been simplified via representation learning. Recent work in this direction use simple feed forward networks \cite{7955211, Estrada2004} and unsupervised auto-encoder based methods \cite{Wen2018}. However, most works generate feature vectors only for a given task and do not try to generalise it to different tasks.



EEGNet \cite{Lawhern2018} is a recent work that uses a CNN based architecture as a universal encoder for various downstream classification tasks. EEGNet attempts to generalize over four tasks which include ERP based task, P300 Speller task, Movement Related Cortical Potential task and the Sensory Motor Rhythm task. 

We consider the EEGNet as our baseline approach since its concept and approach directly competes with our proposed solution and compare it to our designed architecture, as shown in Section \ref{eval_and_results}. We also consider several other models as reference point to validate our results further.

\section{Datasets}\label{datasets}
In this work, we utilize several publicly accessible datasets. The idea was to pick datasets pertaining to a diverse range of EEG based tasks spread across various modalities like text \cite{SHAH2016102}, audio \cite{Yu:2019:DCC:3309717.3281746}, visual \cite{8432497,Shah:2014:APV:2647868.2654919} and more \cite{shah2017multimodal} so that the encoder network can learn different data patterns. Table \ref{tab:datasets_summary} provides a summary of datasets used.

\subsection{SEED: Emotion Detection} \label{SEED}

SEED is a popular dataset for the task of emotion detection, which contains a 62-channel data from 15 participants (7 male and 8 females) while being stimulated by six chosen emotional movie films. The task is essentially a three class classification amongst positive, negative and neutral emotions. The state of the art for SEED tri-class classification reports an accuracy of 83.99\% using SVM and 86.02\% using Deep Belief Network using handcrafted features of Differential Entropy,  Differential Asymmetry and PSD \cite{seed_original}.

\subsection{MindBigData BMNIST: Digit Recognition}\label{MNIST_Brain}

The BMNIST  dataset\footnote{\url{http://www.mindbigdata.com/opendb/}} contains over 1,200,000 brain signal samples of 2 seconds each, acquired at different frequencies. The stimulus is a digit from \numrange[range-phrase = --]{0}{9}. Brain signals are captured when the participant sees and thinks about the exposed stimuli. The data has been captured using Muse headband consisting of 4 channels. Some EEG signals were also captured on random actions and labelled as \emph{-1}.
The brain signals are captured over a course of two years from a single test subject.
The reported state of the art accuracy for this dataset is 
\textcolor{black}{31.35\% \cite{DEvobird2019}}
for 11-class (\numrange[range-phrase = --]{0}{9} or \emph{-1}) classification and 98\% for binary classification (digit or not).


\begin{table}[t]
\normalsize
\caption{Summary of dataset sizes\label{tab:datasets_summary}}
\centering
\begin{tabular}{|l|r|r|} 
\hline
\textbf{Dataset}         & \multicolumn{1}{l|}{\textbf{\# Instances}} & \multicolumn{1}{l|}{\textbf{\# Time steps}}  \\ 
\hline
BMNIST                   & 40983                                             & 408                                                 \\ 
\hline
SEED                     & 318090                                            & 32                                                  \\ 
\hline
ERN                      & 5440                                              & 200                                                 \\ 
\hline
SMR                      & 5184                                              & 500                                                 \\ 
\hline
ThoughtViz       & 51096                                             & 32                                                  \\
\hline
\end{tabular}
\end{table}

\subsection{ThoughtViz: Object Recognition \label{sec:object_thoughtviz}} 
This dataset\footnote{\url{https://github.com/ptirupat/ThoughtViz}} is acquired \textcolor{black}{from Kumar et al.'s work \cite{Kumar2018}}, as described in ThoughtViz \cite{Tirupattur2018}. It contains EEG recordings of 23 participants who were shown stimuli of three kinds - Characters, Digits and day-to-day Objects. Each set contains 10 items, making a total of 30 stimuli (classes). 
The data is collected using Emotiv EPOC + EEG headset with 14 channels at a sampling rate of 128 Hz with post-processing done using a sliding window of 32 samples with an overlap of 8. We have only used the object data for our experiments since we believe that EEG collected from object images as stimulus will evoke more prominent signals and also because BMNIST dataset already incorporates digit stimuli. The state of the art accuracies for digit, character and object 10-class classification is 72.88\%, 71.18\% and 72.95\% using VGG-16 network architecture \textcolor{black}{\cite{Tirupattur2018}}.

\subsection{Sensory Motor Recognition: Task Identification} 
This dataset\footnote{\url{http://www.bbci.de/competition/iv/}}  belongs to the BCI competition 2008, Set \textcolor{black}{2A} \cite{bci_smr}. Data is collected from 9 subjects, who were asked to imagine 4 types of sensory tasks---left hand movement, right hand movement, foot movement or tongue movement. Data is sampled at 250 Hz and a band pass filter was applied at 0.5 Hz and 100 Hz. The state of the art accuracy for within-subject is  68\% and for cross-subject is 40\%. To maintain consistency in all experiments, we have not segregated SMR test data into within or cross-subject. 


\subsection{Feedback Error Related Negativity: Error Detection}\label{ERN}
For the Error Detection task, \textcolor{black}{56 channel EEG data from 26 healthy subjects (13 male) was collected in Margaux et al.'s work \cite{ern_kaggle} and was used in Kaggle's BCI Challenge\footnote{\url{https://www.kaggle.com/c/inria-bci-challenge/}}}. A P300 \cite{P300-review-POLICH20072128} speller is used which is a communication device that allows a subject to spell out text by focusing on each character located at some position on a \textcolor{black}{$6\times6$} grid. While the subject focuses on a character, random flashing of rows and columns on this grid causes an ERP and alters the EEG pattern which is detected. The aim is to determine if the selected item is correct or incorrect by analyzing the brain signals after the subject received feedback. The state of the art AUC reported for this dataset \cite{ERNLeaderboard} using handcrafted features is 0.872, while EEGNet baseline achieved an AUC of 0.8.


\section{Methodology}\label{methodology}
\begin{figure}[h]

    \includegraphics[width = 9.7cm]{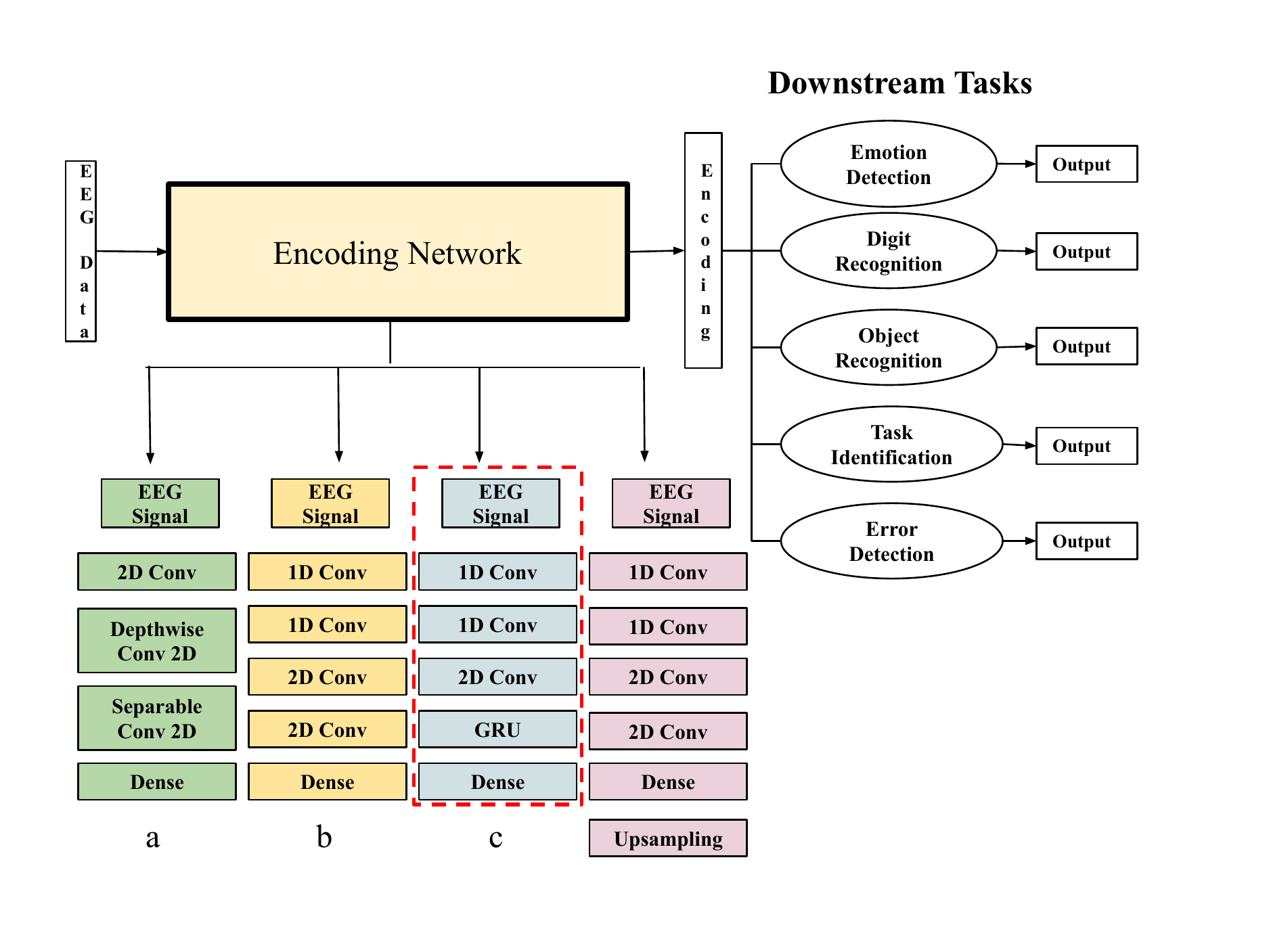}
    \caption{Universal Encoding Architecture a) EEGNet Architecture, b) CNN Architecture, c) GRU Architecture, d) Autoencoder Architecture}
    \label{fig:arch}
\end{figure}

We develop a 
generic architecture for generating encodings for all kinds of EEG data and understand how the learnt representations perform in comparison to existing works on various downstream tasks. We implemented CNN-based, GRU-based and Autoencoder-based networks, which are discussed in the sections below and shown in Figure \ref{fig:arch}. \textcolor{black}{The encoding obtained from network is forwarded to downstream classification task and their performances were analyzed to find the best encoding architecture.} The results obtained over all datasets using each encoding network are evaluated and compared in Table \ref{tab:consolidated_res}. 
The train-test-validation split is kept same as dictated by the respective dataset sources. If it is not specified, we reserve $25\%$ of data for testing and $25\%$ of train data for validation. All models were trained with maximum epochs of 100, and best model was picked based on validation accuracy. The experiments were performed on a system with 50 GB RAM, 16 CPU Cores and 3 NVIDIA GTX 1080Ti GPU.

\vspace{-0.7cm}

\subsection{CNN-based Networks}
CNNs have shown success in capturing spatial dependencies in the input using convolution operations and filters. CNNs utilize parameter sharing and local connectivity to optimize the number of parameters. We have implemented EEGNet \cite{Lawhern2018} architecture. We also replicated the 4 CNN based network using combination of two 1D and two 2D convolutional layers inspired by \textcolor{black}{by Tirupattur et al.} \cite{Tirupattur2018} along with batch normalization and max pooling layers. The convolutional layers consist of (filter, kernels) = (32, (1, 4)), (32, (14, 1)), (50, (4, 25)) and (100, (50, 2)) activated using ReLU and flattened at the output. We use Adam Optimizer with a learning rate of $0.001$, a batch size of $128$ and trained for $100$ epochs using a categorical cross-entropy loss. The fully connected classifier uses two dense layers of size 256 and num\_classes respectively with a dropout having probability 0.5 and ReLU activation for hidden layer and softmax activation for output layer. Public implementation of EEGNet \cite{Lawhern2018} as provided by the authors was used for experiments. 

\begin{figure}[ht]
    \centering
    \includegraphics[width=8.6cm,height=8.6cm,keepaspectratio]{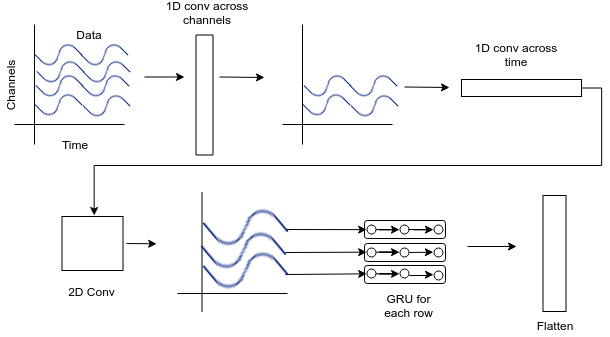}
    \caption{Proposed GRU architecture}
    \label{fig:ourarch}
\end{figure}

\subsection{GRU-based Network}
Since EEG is a time-series data, Recurrent Neural Networks (RNNs) are expected to perform better. They are able to feed the output back into the network. However, they suffer from the vanishing gradient problem. To solve the same the Gated Recurrent Unit (GRU) \cite{learning-rnn-encoder-cho2014, properties-neural-machine-translation-cho2014} was proposed, which uses the update and reset gates. The update gate helps the model decide how much of the past information needs to be passed on to the future iterations, while the reset gate determines how much of the past information is to be forgotten. We have developed a GRU-based network, as shown in Figure \ref{fig:ourarch} where two 1D convolution layers followed by a 2D convolutional layer are used following which the GRU layer is added. The convolutional layers consist of (filter, kernels) = (32, (1, 4)), (32, (14, 1)) and Depthwise 2D Convolution (50, (4, 25)). A GRU sequence of 30 GRU nodes was used for each output channel from the convolutional part. The final encoding is generated by a dense layer. We use Adam Optimizer with a learning rate of $0.001$ and a batch size of $64$ for the training.

\begin{table*}[ht]
\normalsize
\caption{Results on various datasets for all models}
\label{tab:seed}
\label{tab:ERN}
\label{tab:bmnist_2}
\label{tab:bmnist_11}
\label{tab:thoughtviz_image}
\label{tab:smr_result}
\label{tab:consolidated_res}
\centering
\begin{tabular}{|l|r|r|r|r|r|r|r|r|r|r|r|r|} 
\hline
                  & \multicolumn{2}{c|}{ERN}                           & \multicolumn{2}{c|}{SMR}                           & \multicolumn{2}{c|}{BMNIST}                        & \multicolumn{2}{c|}{BMNIST\_2}                     & \multicolumn{2}{c|}{SEED}                          & \multicolumn{2}{c|}{ThoughtViz}                     \\ 
\hline
                  & \multicolumn{1}{c|}{Acc} & \multicolumn{1}{c|}{F1} & \multicolumn{1}{c|}{Acc} & \multicolumn{1}{c|}{F1} & \multicolumn{1}{c|}{Acc} & \multicolumn{1}{c|}{F1} & \multicolumn{1}{c|}{Acc} & \multicolumn{1}{c|}{F1} & \multicolumn{1}{c|}{Acc} & \multicolumn{1}{c|}{F1} & \multicolumn{1}{c|}{Acc} & \multicolumn{1}{c|}{F1}  \\ 
\hline
EEGNet \cite{Lawhern2018}          & 0.712                    & 0.421                   & 0.549                    & 0.528                   & 0.329                    & 0.138                   & 0.942                    & 0.926                   & 0.547                    & 0.545                   & 0.251                    & 0.233                    \\ 
\hline
Autoencoder KNN           & 0.665                    & 0.515                   & 0.260                    & 0.210                   & 0.276                    & 0.056                   & 0.846                    & 0.785                   & 0.393                    & 0.381                   & 0.419                    & 0.424                    \\ 
\hline
Autoencoder RF            & 0.630                    & 0.529                   & 0.243                    & 0.137                   & 0.275                    & 0.042                   & 0.857                    & 0.817                   & 0.365                    & 0.305                   & 0.651                    & 0.702                    \\ 
\hline
4 CNN Network               & 0.711                    & 0.420                   & 0.385                    & 0.383                   & 0.352                    & 0.152                   & 0.994                    & 0.993                   & 0.648                    & 0.644                   & 0.740                    & 0.740                    \\ 
\hline
\textbf{GRU Network} & \textbf{0.714}           & \textbf{0.433}          & \textbf{0.333}           & \textbf{0.296}          & \textbf{0.338}           & \textbf{0.160}          & \textbf{0.993}           & \textbf{0.991}          & \textbf{0.744}           & \textbf{0.744}          & \textbf{0.774}           & \textbf{0.774}           \\
\hline
\end{tabular}
\end{table*}

\subsection{Autoencoder Network}
An autoencoder is an unsupervised technique used to learn data encoding. It has two parts---the encoder compresses data and reduces dimensionality to generate an \textit{encoding}, while the decoder tries to regenerate the original input using the encoding. For our experiment, we use the encoding learnt by the autoencoder and redirecting it to a suitable classifier like random forest or $K$ nearest neighbours. The encoder uses convolutional layers from 4CNN architecture (two 1D and two 2D convolutions; ReLU activations; (filter, kernels) = (32, (1, 4)), (32, (14, 1)), (50, (4, 25)) and (100, (50, 2))) to extract feature representations which are max-pooled and normalized using Batch Normalization. The extracted representations are then flattened together and connected to a Dense layer with the embedding size of 128, which marks the end of encoder. For decoding phase, a stack of two dense layers (equal to the flattened output and input data's shape respectively) with ReLU activation are used and then this raw output is used for calculating loss. We use AdaDelta optimizer with a learning rate of $0.001$, batch size of $128$ and binary cross-entropy as loss function. KNN and Random forest classifier uses sklearn's default hyperparameters, viz 5 neighbors with uniform weighting for KNN and 100 estimators with no restriction on maximum tree depth for random forest classifier.



\section{Results and Discussion}\label{eval_and_results}

We evaluated the performance of all architectures (Section \ref{methodology}) on the five downstream tasks mentioned in Section \ref{datasets} and compared their performances, as shown in Table \ref{tab:consolidated_res}.

Autoencoder models are not the top performing models in any of the downstream tasks. For SEED
, GRU-based model significantly outperforms all other models. In the ERN 
and BMNIST (11 class) 
tasks, GRU-based shows similar performance to 4 CNN and EEGNet. In BMNIST (2 class) 
and ThoughtViz 
, GRU-based is the best performing model. The 4 CNN network gives a similar performance and both significantly outperform EEGNet in these tasks. In ThoughtViz EEGNet has an accuracy of 25\%, significantly lesser than accuracy from all other models.

We observe that GRU-based network is, for most tasks, the best performing model, with 4 CNN giving similar performance in some cases. This shows the importance of using temporal features to make accurate predictions on EEG continuous signal data.

In the SMR task (Table \ref{tab:smr_result}) however, both 4 CNN and GRU-based models are significantly outperformed by EEGNet.
From Table \ref{tab:datasets_summary} we see that SMR and ERN are the two smallest datasets. 
The SMR dataset is an Oscillatory BCI, while ERN is Event Related Potential (ERP) BCI \cite{EEG-task-types-Lotte}. In Event Related Potential (ERP) BCI tasks, the EEG response to a known external stimulus is detected, while in Oscillatory BCI tasks, signals are used from specific frequency bands for controlling the interface. ERP tasks can be modelled by Machine learning more efficiently than Oscillatory BCI tasks \cite{p300-challenges-trends-fazel}. This is because ERPs have high amplitude and low frequency waveforms. Due to more dependency on external stimuli, these are more robust across subjects. On the other hand, Oscillatory tasks use brain signals for external control. They are harder to train, due to lower signal-to-noise ratio \cite{motor-imagery-bci-pfurtscheller} and there is more variation across subjects. The signals in Oscillatory BCIs are asynchronous \cite{motor-imagery-bci-pfurtscheller}, which can be a limitation  for GRUs, as they are designed for temporal (Sequential) detection and work across the frequencies.

We also observe that GRU-based network significantly outperform previous state of the art results on BMNIST and ThoughViz. On SEED, despite being the best performing model, it still lags behind the results achieved using handcrafted features \cite{seed_original}.

\balance

\section{Conclusion and Future Work}\label{concl_future}
Through this work, we develop a common encoding architecture for a variety of tasks. We can claim that using deep learning based approaches, features significant to each of the seemingly different tasks can be extracted and used for classification. We have experimented with local analysis using CNN, temporal analysis using GRUs and unsupervised analysis using Autoencoders. We achieved the best performance using the GRU-based model where we were also able to beat the best-reported performances on each of the downstream tasks, with SMR task being the only exception due to its oscillatory nature. This implies how temporal features extracted from a GRU are successfully able to represent the diversity in EEG data. Through such experiments, we have introduced a variety of methods to study EEG data using universal encodings and have in-turn contributed to research in the field of BCI and neuroscience.

As part of future work, attention-based mechanisms can be employed to further enhance the performance of the GRU-based network. Oscillatory BCI tasks are a limitation for the GRU-based architecture, which should be dealt with. The auto-encoders can be further improved by guiding the decoder using encoder (e.g using skip connections). Further, since all these tasks involve EEG signal data, Multi-Task Learning can be employed to learn collectively from the various tasks.



\section*{Acknowledgement}\label{acknowledgement}
Rajiv Ratn Shah is partly supported by the Infosys Center of AI, IIIT Delhi and ECRA Grant by SERB, Govt. of India.

\newpage
\bibliographystyle{IEEEtran}
\bibliography{references}

\end{document}